# EFFECTS OF ARRIVAL TYPE AND DEGREE OF SATURATION ON QUEUE LENGTH ESTIMATION AT SIGNALIZED INTERSECTIONS

**Behnoush Garshasebi**
Department of Civil and Environmental Engineering
Room B106, Newmark Civil Engineering Laboratory
University of Illinois at Urbana-Champaign, Urbana, IL 61801 USA
Tel: 217-979-0365; Email: garshas2@llinois.edu

**Rahim (Ray) F. Benekohal**
Department of Civil and Environmental Engineering
1213, Newmark Civil Engineering Laboratory
University of Illinois at Urbana-Champaign, Urbana, IL 61801 USA
Tel: 217-244-6288, Fax: 217-333-1924 Email: rbenekoh@illinois.edu

**ABSTRACT**

Purpose of this study is evaluation of the relationship between different arrival types and degree of saturation (X) with overestimations of HCM 2010 procedure for estimating the back of queue within a study area. Further analysis is performed to establish the relationship between queue length and delay and also between each of them individually and X in cases with overestimation. The analyses are based on the $50^{th}$ percentile queue lengths for data collected at four signalized intersections along a corridor in 4 time periods (off peak period and AM, Noon and PM peak periods).

Based on the statistical test results, arrival type did not play a role in overestimations. However, there is a significant relationship between the overestimations on minor and major street and different ranges of X. On minor streets, about 59% of the overestimations are at X values less than half; while near 23% of the overestimations are at oversaturation condition with X values greater than 1.

The relationship between amount of overestimations and degree of saturation should be established based on the numerical amount of overestimations versus X values rather than the relative amounts; since the statistical comparison between the relative amount of overestimations and X values, resulted in a wrong idea of the real world condition.

There was a significant correlation between field queue and delay data of the cases with overestimated queue length in all cases on major and minor streets. Also, field queue is correlated to X, in all cases on minor and major streets.

*Keywords*: Queue length estimation, HCM, queue, delay, signalized intersections, arrival type, degree of saturation, queue overestimation, delay and queue in field data, signal coordination.

# INTRODUCTION

Queue length is one of the principal performance measures for signalized intersections and may affect signal timing and geometric design decisions. Queue length at signalized intersections is estimated for dimensioning of the left turn lanes and evaluation of the adequacy of storage space (1). There are multiple methods for queue length estimation at signalized intersections. A great majority of the existing methods are based on two models: the input-output model (2, 3) and the shockwave model (4). These models were applied into methods, to estimate the average (or steady-state) queue length over a relatively long time interval; e.g., 15 min or 1 hour (5). Methods for estimating average queue lengths at signalized intersection are well established (6). Vitoria et al. (7) compared queue length estimation methods from SIDRA, HCM2000, TRANSYT- 7F, SOAP, NCHRP 279 Guidelines, SIGNAL 97, NETSIM and Oppenlander's Method. In their study, Highway Capacity Manual (HCM) 2000 queue model provided more accurate estimations than the other methods and was recognized as a standard procedure for queue length estimation. HCM 2010 (8) presented an improved back-of-queue calculation model for signalized intersection compared to HCM 2000 method (9), by taking into account the effects of initial queue. Some studies evaluated the HCM queue length model for interchange ramp terminals and all-way stop controlled intersections (10, 11).

Error in queue length estimation, may lead to inappropriate left-turn lane design which may cause additional delay at the intersections; and left-turn lanes entrance blockage by the queued through traffic (13). Generally, overdesigning for an intersection should be avoided due to negative impacts to pedestrians due to wider street crossings, the potential for speeding, land use impacts, and cost (14). Liu (12) used filed data to assess the accuracy of HCM2010 back-of-queue estimation model. In that study the $50^{th}$ percentile queue length estimates from the HCM2010 model and field data were compared; which resulted in significant discrepancies is 52% of the cases, of which in 93% of the cases were overestimation of the queue length. The above mentioned study (12) did not investigate the relationship between the overestimations and some traffic parameter such as degree of saturation, traffic volume, arrival type and other factors that could play a role.

This study investigate whether the queue length overestimations of HCM 2010 as discussed in Liu et al. (12) depend on arrival type and/or degree of saturation; and if there is a relationship between field queue and associated delays on minor and major streets; and also between these variables and degree of saturation.

It should be noted that, the HCM 2016 queue length estimation method is the same as the HCM2010 method. So, the findings are applicable to HCM 2016 as well.

# DATA COLLECTION

## Description of the Study Area

The study area of interest consists of six intersections along the Neil Streets corridor, Champaign, IL (Figure 1). At the time of data collection, the six intersections on Neil Streets were operating as time-based coordinated signals and provided progression for northbound and southbound traffic (the major street). The traffic pattern on Neil Streets is such that in the morning it has higher volume going northbound (toward downtown Champaign), but in the afternoon it is the southbound that has higher volume. Four of the crossing streets that create typical four-legged intersections are Stadium Drive, Kirby Avenue, St. Mary's Road and Windsor Road. Heavy volume direction in the morning is eastbound towards the campus of University of Illinois at Urbana-Champaign, and westbound in the afternoon. This study uses the data collected from the four typical four-legged intersections. Data for atypical intersections (the T-intersection of Devonshire drive and Neil St as well as Knollwood Drive and Neil St which is only a driveway for a commercial plaza) are not used.

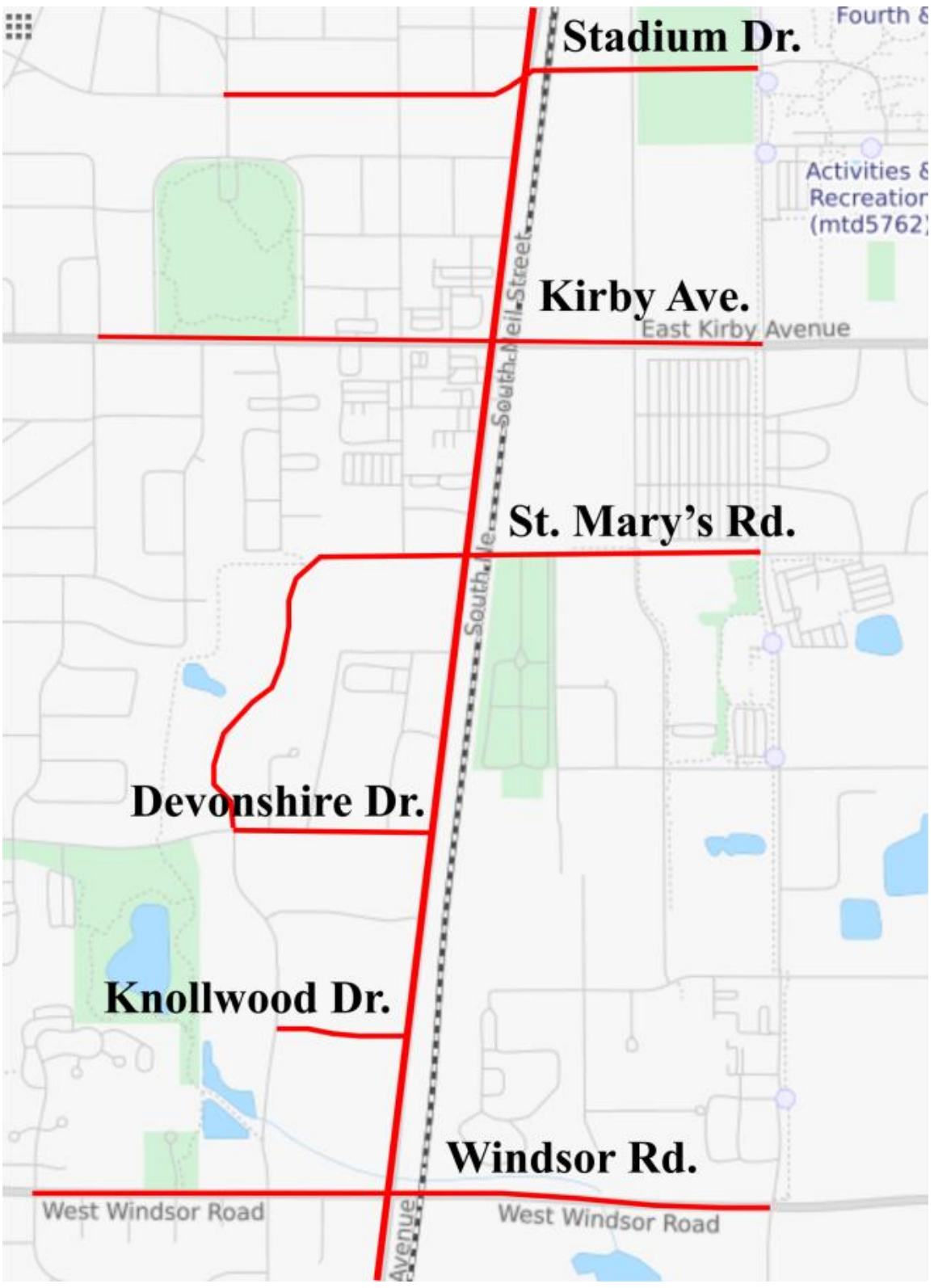

**FIGURE 1 Six study intersections along the Neil Street corridor, Champaign, IL (15)**

## Field Data Collection Methodology

Data collection was conducted by recording videos of the traffic conditions during multiple time periods in a day to get data during morning peak (7:10 AM - 8:40 AM), noon peak (10:40 AM - 1:15 PM), and pm peak (4:40 PM - 6:00 PM). The field traffic data for this study were collected between October 29 and December 11.  The dates and days corresponding to data reduction at each intersection are shown in table 1.

**TABLE 1  Dates and Day of Data Reduction**

| Intersection | Date | Day |
|---|---|---|
| Neil Street and Stadium Drive | November 7, 2013 | Thursday |
| Neil Street and Kirby Avenue | November 13, 2013 | Wednesday |
| Neil Street and St. Mary's Road | November 20, 2013 | Wednesday |
| Neil Street and Devonshire Drive | October 29, 2013 | Tuesday |
| Neil Street and Knollwood Drive | November 12, 2013 | Tuesday |
| Neil Street and Windsor Drive | November 5, 2013 | Tuesday |

## Data Reduction

The following traffic characteristics data were extracted from the video: peak hour, hourly volume, saturation flow rate, signal timing, arrival type, and truck volume. After the peak hours were determined, stopped delay and queue length was obtained for one hour for each time period. Detailed description of data reduction is presented in a previous study (12) and a brief description of stopped delay computation is given below.

## Stopped Delay Data Collection

The field stopped delay was calculated by using the data reduced from the videos. The field measurement technique for intersection control delay given in Chapter 31 of HCM 2010 (8) was adapted to calculate stopped delay. The Stopped delay was calculated for a lane or a lane group of each approach of the intersections. The stopped delay was used (as traditionally it has been used) in the comparison instead of control delay because it is directly computed from field data and in not affected by adjustments that are applied to get the control delay. The procedure requires identifying the approach speed during each study period. The speed limit of each approach individually was assumed to be its approach speed for each intersection. The data used is for a period of time that was essentially equal to 1 hour for each peak hour and the off-peak hour. The count interval of 15 seconds was selected for stopped delay calculation because it is an integral divisor of the duration of survey period (1 hour) as required by the HCM 2010 procedure (8).

## Method of Queue Length Estimation

*HCM 2010 Back-of-Queue Procedure*

Liu (12) used HCS 2010 version 6.70 to perform the 50$^{th}$ percentile back-of-queue calculation procedure provided by Highway Capacity Manual 2010 (8). Inputs for traffic characteristics in Highway Capacity Software (16) runs were also obtained from the field data. More detailed calculation process for back-of-queue can be found in HCM 2010 (8).

## Results from the 50$^{th}$ Percentile Queue Length Comparison

Liu (12) statistically compared the 50$^{th}$ percentile queue length estimations from HCM 2010 and the 50$^{th}$ percentile queue length from the field. Table 2 shows the summary of comparison results. The single-sample Wilcoxon signed-rank test with 90% confidence level resulted in p-values smaller than 0.1 for 27 out of 52 cases (52%), indicating significant discrepancies between the HCM 2010 queue length estimates and field data. When there was a significant discrepancy, in 93% of the cases HCM significantly overestimated the queue length which is an overwhelming great majority of cases.

About 1/3 of the cases on major streets and 3/4 of the cases on minor streets had significant differences in comparison with the filed data. Distribution of such significant overestimations frequency over four typical intersections along the corridor of interest, are represented in Figure 2. When there was a significant discrepancy, HCM overestimated in 89% of the major street cases and in 94% of the minor street cases.

**TABLE 2 The 50th Percentile Queue Length Comparison Summaries (12)**

| Categories | Number of Cases | Percentage[1] | Range of (HCM-Field)/ Field %[2] | Average Discrepancy (%) |
|---|---|---|---|---|
| **Overall** | | | | |
| **Total** | 52 | - [3] | - [3] | - [3] |
| **Significant Discrepancies** | 27 | 52% | (-42)-137% | - |
| **Significant Overestimation** | 25 | 93% | 14-137% | 52% |
| **Significant Underestimation** | 2 | 7% | (-42)-(-20)% | (-31)% |
| **Major Street (NB[4]/ SB[4])** | | | | |
| **Total** | 28 | - | - | - |
| **Significant Discrepancies** | 9 | 32% | (-42)-120% | - |
| **Significant Overestimation** | 8 | 89% | 17-120% | 66% |
| **Significant Underestimation** | 1 | 11% | (-42)% | (-42)% |
| **Minor Street (EB[4]/ WB[4])** | | | | |
| **Total** | 24 | - | - | - |
| **Significant Discrepancies** | 18 | 75% | (-20)-137% | - |
| **Significant Overestimation** | 17 | 94% | 14-137% | 44% |
| **Significant Underestimation** | 1 | 6% | (-20)% | (-20)% |

NOTE: [1] In each Percentage column, the percentage correspond sequentially to $\frac{\text{Significant Disrepancies}}{\text{Total}}$ %, $\frac{\text{Significant Overestimation}}{\text{Significant Discrepancies}}$ %, and $\frac{\text{Significant Underestimation}}{\text{Significant Discrepancies}}$ %.

[2] (HCM-Field)/ Field %, which computes the relative amount of discrepancy, for eastbound PM and westbound AM at Neil Street and Stadium Drive are unavailable, because the field queue lengths of these two cases are 0 veh/ln.

[3] Data not applicable

[4] NB, SB, EB, and WB are the abbreviations of northbound, southbound, eastbound and westbound, respectively. Same abbreviations will be used in the following tables and figures.

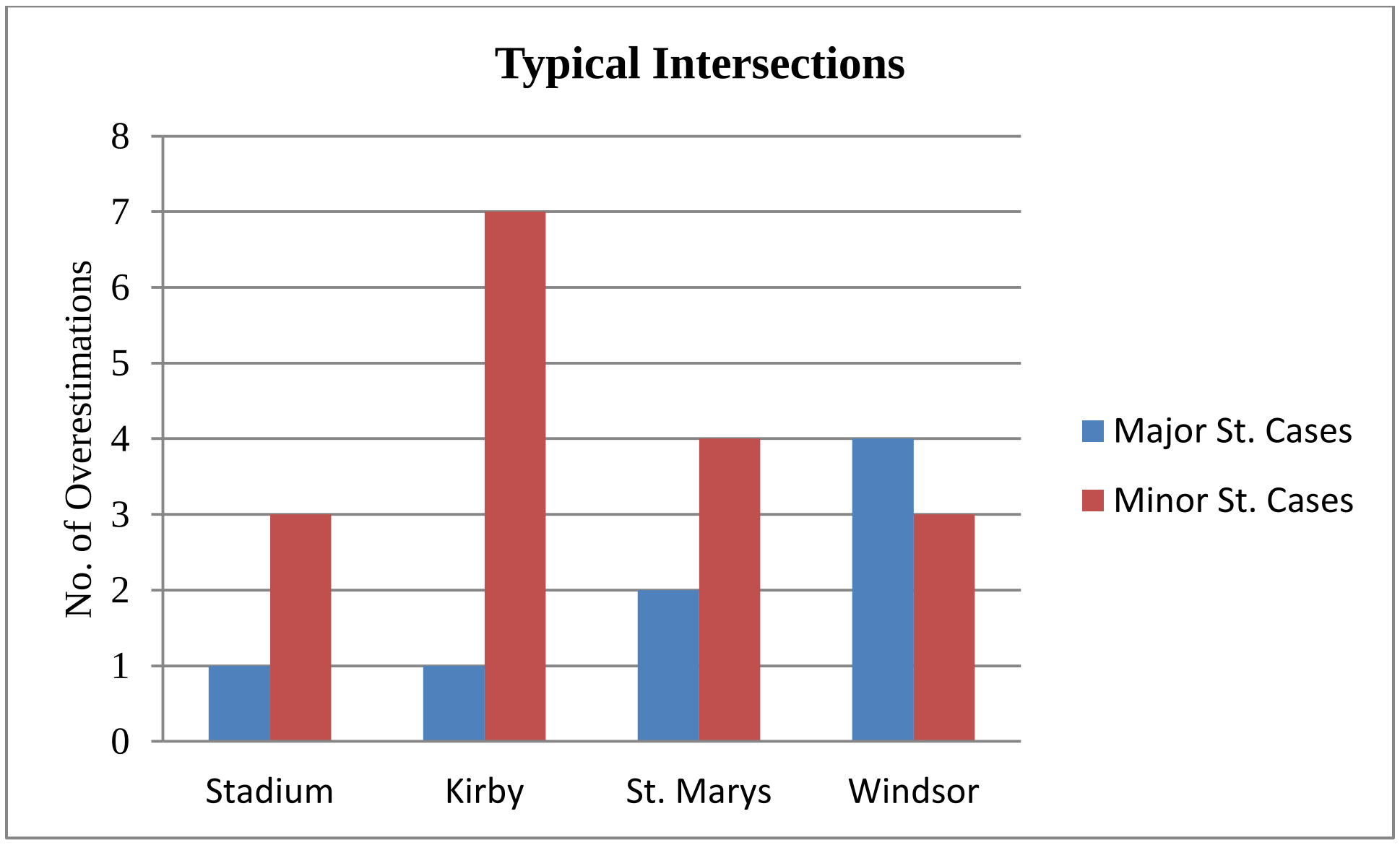

**FIGURE 2 Overestimation on the four typical intersections on Neil street corridor**

The above mentioned study (12) did not investigate the relationship between the overestimations and some traffic parameter such as degree of saturation, traffic volume, arrival type and other factors that could play a role. This study explores these relationships to get some insightful information.

## DATA ANALYSIS AND DISCUSSION OF FINDINGS

This study conducts statistical analysis to explore if the discrepancies (overestimations) in queue length estimations by HCM 2010 procedure as presented in Liu (12) depend on arrival type or degree of saturation. Then a further statistical analysis (17) was conducted to analysze if there is a relationship between field queue, and associated delays on minor and major streets, and also between these variables and degree of saturation. As previously mentioned, the overwhelming majority of the significant discrepancies was overestimation, only cases with significant overestimation, are used for the purpose of statistical tests and further analysis in this study.

### Relationship between Arrival Type and Queue Length Overestimation in HCM

The arrival type (AT) in the Highway Capacity Manual (HCM) is a measure of quality of progression from one signalized intersection to the next along a coordinated corridor. One way of determining AT is using the percentage of vehicles arriving during the green indication, when they arrive during the green interval, and the density of the arriving platoon (18). Another way of determining AT is observing the time during a cycle that the platoon arrives. In our study area the AT for the four typical coordinated intersections was determined to be arrival types 2, 3, or 4 by watching the video tapes of the intersections. The detailed information about these procedures is given in the previous study (12).

To explore the relationship between different arrival types and the overestimations of queue length on minor and major streets, they are further categorized into two groups; one group includes arrival type 4 and the other one includes arrival types 2 and 3 combined (Individually AT 2 and 3 had small numbers). Table 3 shows the arrival types of the minor and major street approaches with overestimated queue lengths.

A Chi-square test of goodness-of-fit was performed to find any association between the arrival types and the overestimations on minor and major approaches with a significance level of 0.1. By the given evidence, we fail to reject the null hypothesis, indicating that there is no significant association between distribution of overestimations on major and minor streets and different arrival types, $\chi^2$ (1, N = 25) = 0.618, p > 0.1.

This indicated that the AT (the favorable AT 4 or unfavorable AT 2 or random AT3) did not play a significant role in distribution of queue length overestimations on minor or major street approaches.

**TABLE 3 Arrival types of minor and major street approaches with overestimated cases**

| Intersection | Time Period | NB | SB | EB | WB |
|---|---|---|---|---|---|
| | | THRU | THRU | THRU | THRU |
| Neil St & Stadium Dr | AM Peak | *N.A* | *N.A* | **4.0** | **3.0** |
| | Off Peak | *N.A* | *N.A* | *N.A* | *N.A* |
| | Noon Peak | **4.0** | *N.A* | *N.A* | *N.A* |
| | PM Peak | *N.A* | *N.A* | **4.0** | *N.A* |
| Neil St & Kirby Ave | AM Peak | *N.A* | *N.A* | **4.0** | *N.A* |
| | Off Peak | *N.A* | *N.A* | **4.0** | **4.0** |
| | Noon Peak | **2.0** | *N.A* | **3.0** | **4.0** |
| | PM Peak | *N.A* | *N.A* | **4.0** | **2.0** |
| Neil St & St Mary's Rd | AM Peak | **4.0** | *N.A* | **4.0** | **3.0** |
| | Off Peak | *N.A* | *N.A* | *N.A* | *N.A* |
| | Noon Peak | **3.0** | *N.A* | *N.A* | *N.A* |
| | PM Peak | *N.A* | *N.A* | **2.0** | ***3.0*** |
| Neil St & Windsor Rd | AM Peak | *N.A* | **4.0** | **4.0** | *N.A* |
| | Off Peak | *N.A* | **4.0** | **4.0** | *N.A* |
| | Noon Peak | *N.A* | **4.0** | *N.A* | *N.A* |
| | PM Peak | *N.A* | **4.0** | *N.A* | **3.0** |

## Relationship between Degree of Saturation (v/c ratio) and Queue Length Overestimations in HCM

The degree of saturation, also referred to as v/c ratio, is given the symbol X in intersection analysis and represents the ratio of volume over capacity (14). For a given lane group $i$, $X_i$ is computed using Equation 1:

$$X_i = (\frac{v}{c})_i = \frac{v_i}{s_i(\frac{g_i}{C})} = \frac{v_i C}{s_i g_i} \tag{1}$$

Where

$X_i = (\frac{v}{c})_i$ = ratio for lane group i,
$v_i$=actual or projected demand flow rate for lane group i (veh/h),
$s_i$=saturation flow rate for lane group i (veh/h),
$g_i$=effective green for lane group i (s), and
C= cycle length (s)

In this study, we computed the X values for each approach of the study area by using the data collected from the field (12) as input values. Table 4 shows the X values for the cases with significant overestimation (note that NB/ SB is the major and EB/WB is minor street)

**TABLE 4 Values of X for cases with overestimations**

| Intersection | Time Period | NB | SB | EB | WB |
|---|---|---|---|---|---|
| | | THRU | THRU | THRU | THRU |
| Neil St & Stadium Dr | AM Peak | N.A | N.A | **0.340** | **0.07** |
| | Off Peak | N.A | N.A | N.A | N.A |
| | Noon Peak | **0.738** | N.A | N.A | N.A |
| | PM Peak | N.A | N.A | **0.065** | N.A |
| Neil St & Kirby Ave | AM Peak | N.A | N.A | **1.122** | N.A |
| | Off Peak | N.A | N.A | **0.442** | **0.305** |
| | Noon Peak | **0.978** | N.A | **0.629** | **0.420** |
| | PM Peak | N.A | N.A | **0.920** | **1.086** |
| Neil St & St Mary's Rd | AM Peak | **0.886** | N.A | **0.242** | **0.108** |
| | Off Peak | N.A | N.A | N.A | N.A |
| | Noon Peak | **0.764** | N.A | N.A | N.A |
| | PM Peak | N.A | N.A | **0.175** | **0.225** |
| Neil St & Windsor Rd | AM Peak | N.A | **0.441** | **1.239** | N.A |
| | Off Peak | N.A | **0.560** | **0.501** | N.A |
| | Noon Peak | N.A | **0.724** | N.A | N.A |
| | PM Peak | N.A | **1.156** | N.A | **1.150** |

For a better understanding of what the relationship between the X values and the overestimations on the minor and major streets might be, the X values are categorized into five groups. Table 6 shows the contingency table with regard to the overestimations frequency distribution over different X values. A Fisher's exact test (with a significance level of 0.1) was performed to determine whether there is a significant correlation between overestimations of queue length on minor and major streets and different X values. The Fisher's exact test yielded a p=0.06, which indicates that the overestimations, are related to the X values with a significance level of 90%. Figure 3 also shows the frequency distribution of overestimations on minor and major streets over the different X values.

**TABLE 5 Fisher's exact test contingency table**

| | X: [0.0,0.24] | X: [0.25,0.49] | X: [0.50,0.74] | X: [0.75,0.99] | X: [1.00,1.24] |
|---|---|---|---|---|---|
| **Overestimations on minor St.** | 6 | 4 | 2 | 1 | 4 |
| **Overestimations on major St.** | 0 | 1 | 3 | 3 | 1 |

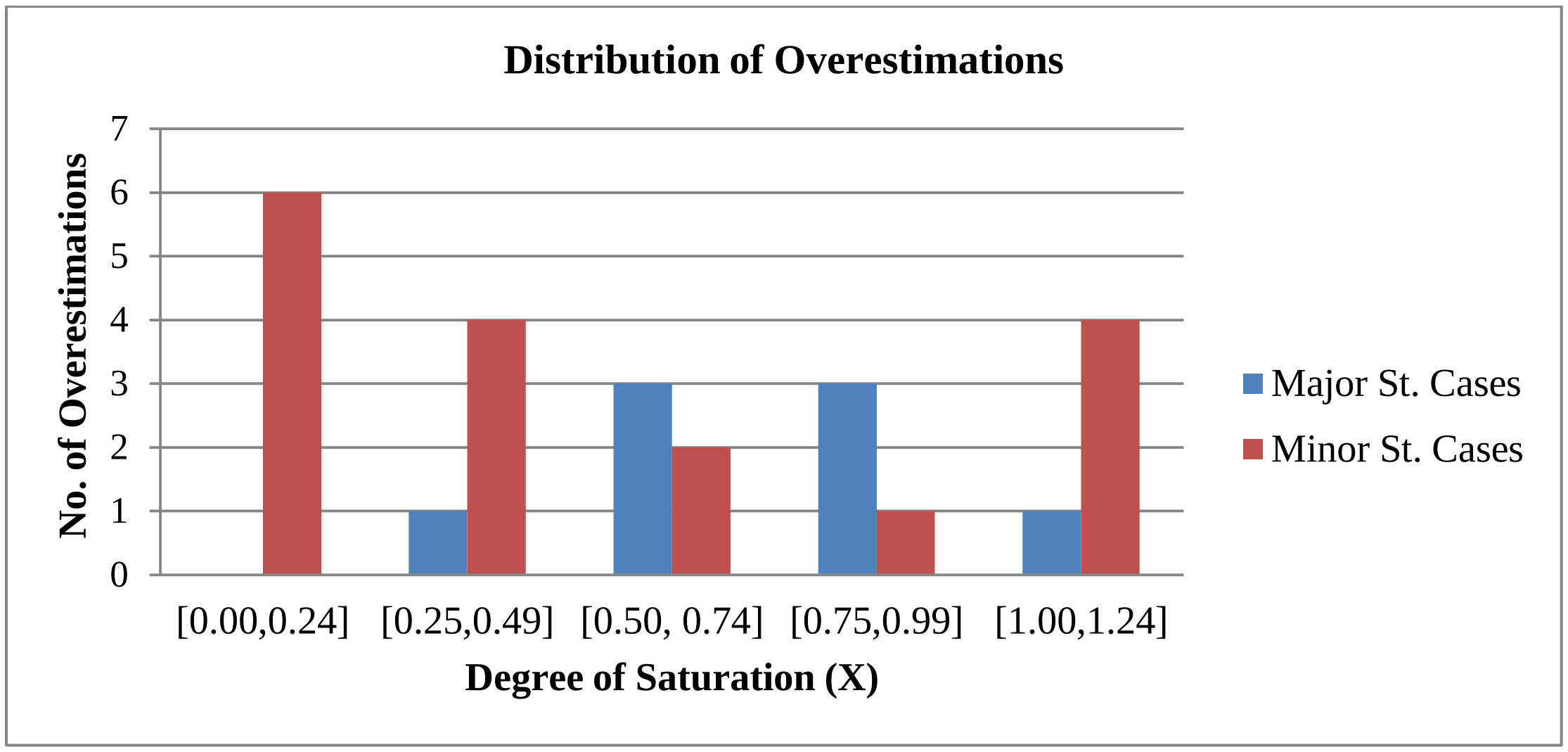

**FIGURE 3 Distribution of overestimations on major and minor streets at different X values**

As shown in Figure 3, more than half (59%) of the overestimations on minor street, are when X<0.49. The overestimations was not limited to the lower X value on minor streets, as 23% of the overestimations happened in oversaturated conditions (X>1).

None of the overestimations on the major street were within the lower X range (X<0.50); while 59% of overestimations on the minor streets fall into this range. The overestimations on the major streets were in medium to high X values. The overestimations on the major street in oversaturation conditions (X>1), included only one observation (12.5% of all overestimations for the major street); the other 87.5% of the overestimations on major street are scattered over the medium to high X range zones (0.44<X<1.00).

**Relationship between Degree of Saturation (X) and Numerical and Relative Amount of Queue Length Overestimation in HCM 2010**

The numerical amount of queue length overestimation (in vehicles) versus X values was analyzed to see if there is a relationship. A Pearson product-moment correlation coefficient (with a significance level of 0.1) was computed to assess the relationship between those two variables. There was a significant positive correlation between the numerical overestimations and X values on for the combined cases of the minor and major streets (r=0.603, p<0.1). The scatter plot summarizes the results, such that the numerical amount of overestimations and X, except within the combined cases on major and minor streets, are also significantly correlated on minor street cases (r= 0.622, p<0.1).

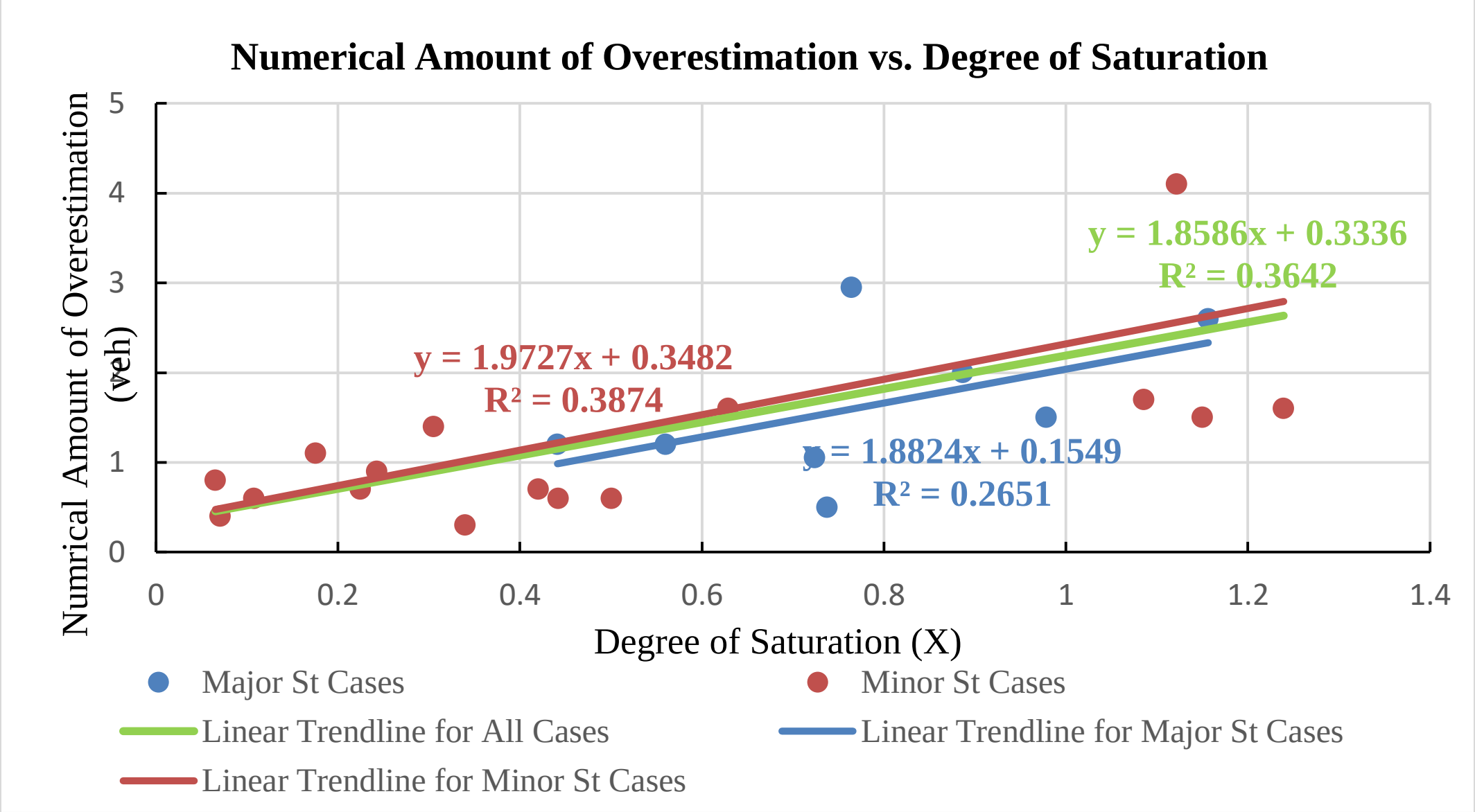

**FIGURE 4 Numerical amount of overestimation (veh) at different X values**

Figure 5 shows the graphical comparison between the relative amount of overestimations ((HCM Queue Length-Field Length)/Field Queue Length %) on minor and major streets versus the X values. The graph shows that the overall regression line through combined cases on minor and major streets shows that as X increases, the relative amount of overestimations also increases, though slightly. Similar trend is true for the minor street cases. However, as the X increases, percentage of overestimation on the major street decreases. It implies that the analysis based on the combined minor and major street cases could generate a wrong idea about the real world trend. The relationship between the relative amount of overestimations and the X values should be performed separately for the cases on minor and major streets. The trend for the major street cases is opposite of the trend for the minor street cases. For the major street the relative error decreases as X increases, as expected. This trend supports the trend that was observed on Figure 4 for the major street cases in which the number of vehicles was considered.

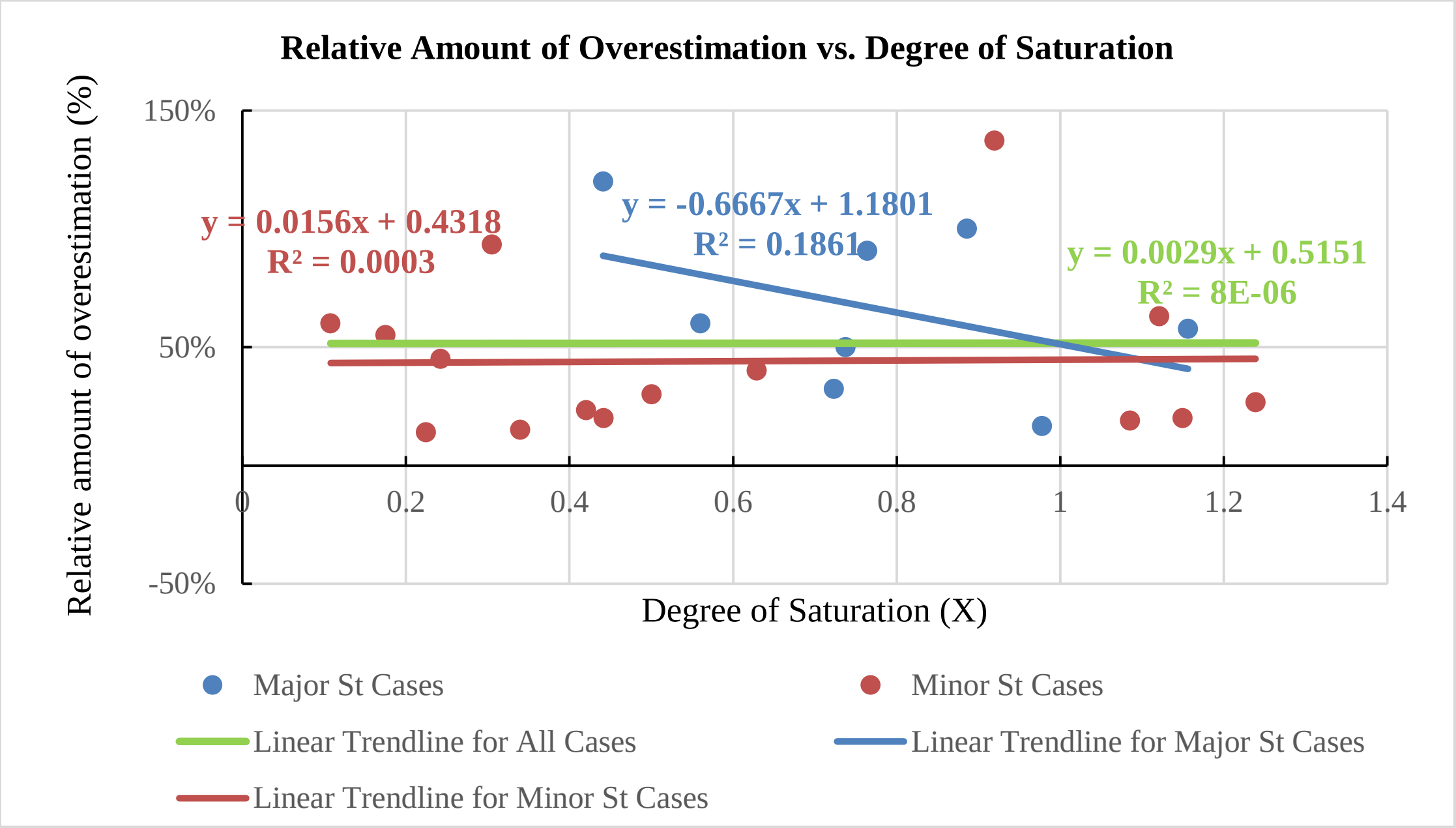

**FIGURE 5 Relative Amounts of Overestimations versus Degree o Saturation (X)**

**Relationship between the Field Queue Length Overestimation, Delay and X**

The delay data collected from the field are statistically compared to the field queue lengths of the cases with overestimations on minor and major streets. A Pearson product-moment correlation coefficient (with a significance level of 0.1) was computed and it resulted in a significant correlation between queue and delay of combined cases on major and minor streets (r=4882, p<0.1). Further analysis was performed to assess the potential relationships between queue and delay on major and minor street cases, separately. There is a significant correlation between queue and delay on minor street (r= 0.4517, p<0.1). Performing the test on major street cases data, the null hypothesis was rejected (r=0.9627, P<0.1), which shows a significant correlation between overestimated queue length and delay on major street.

On the other hand, some potential trends might exist among queue data and degree of saturation at signalized intersections and for the delay data likewise. For revealing these probable correlations, the relationship between queue and degree of saturation (X) was statistically analyzed. Figure 6 shows the scatter plot associated with this analysis. Testing the data, we concluded that there is a strong relationship between queue and X for the combined minor and major street cases (r= 0.7576, p< 0.1), and individually on minor street cases (r= 0.8684, p< 0.1) and major street cases (r=0.6835, p<0.1), (Figure 7)

In order to examine the relationship between delay data and X values, the data point were plotted and regression lines were fitted (Figure 8).

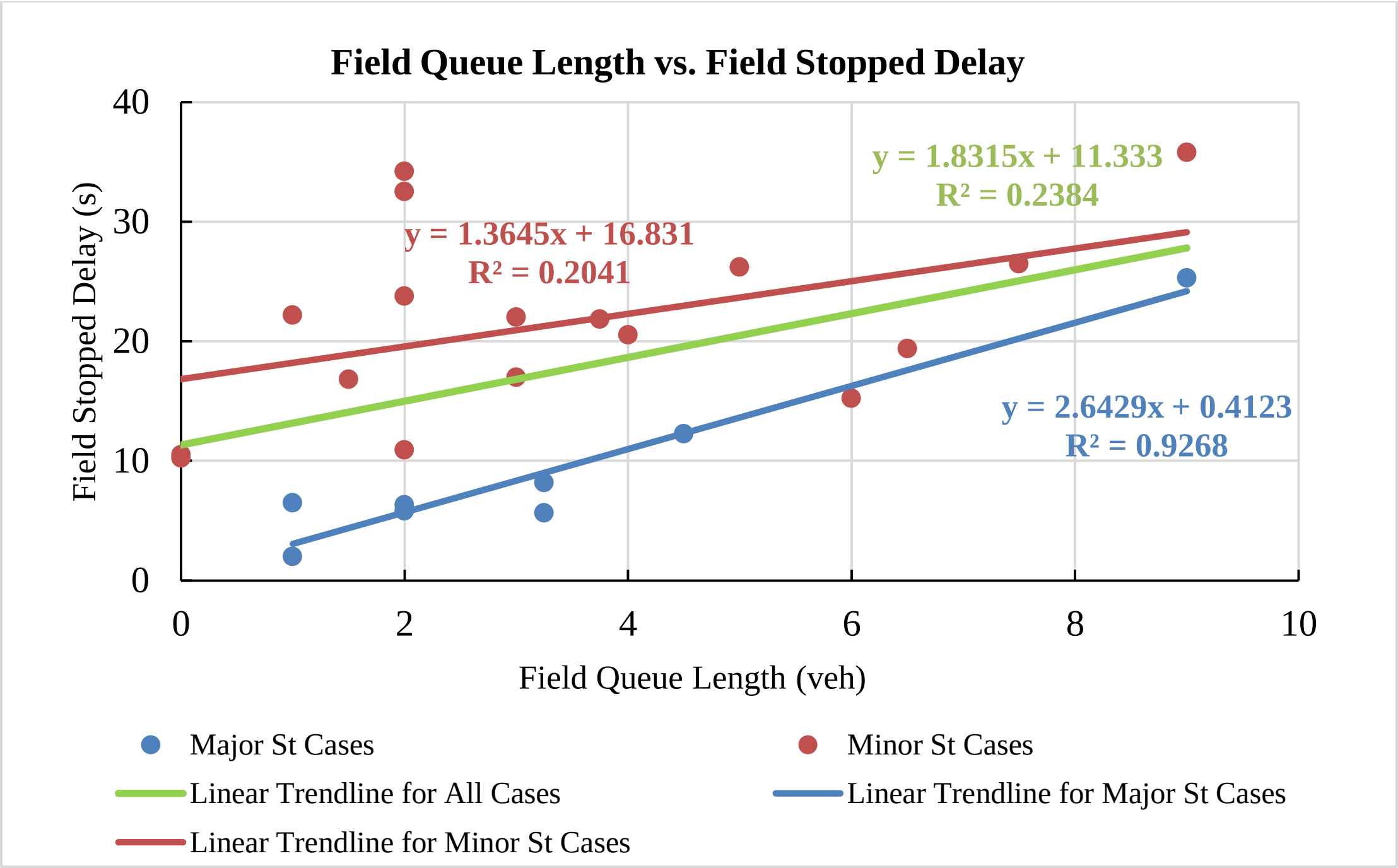

**FIGURE 6 Field Queue Length versus Field Stopped Delay Scatter Plot**

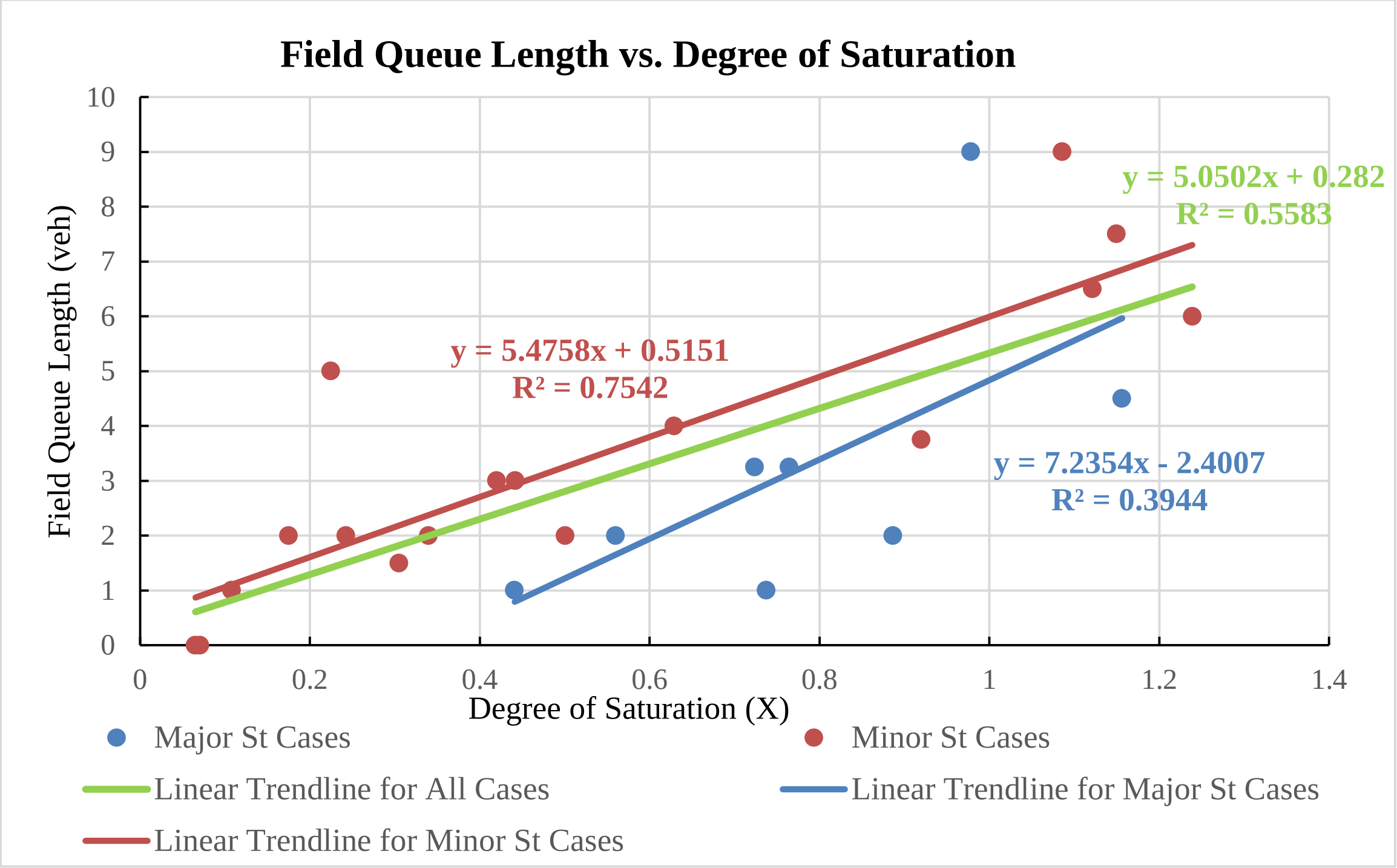

**FIGURE 7 Field Queue Lengths versus Degree of Saturation**

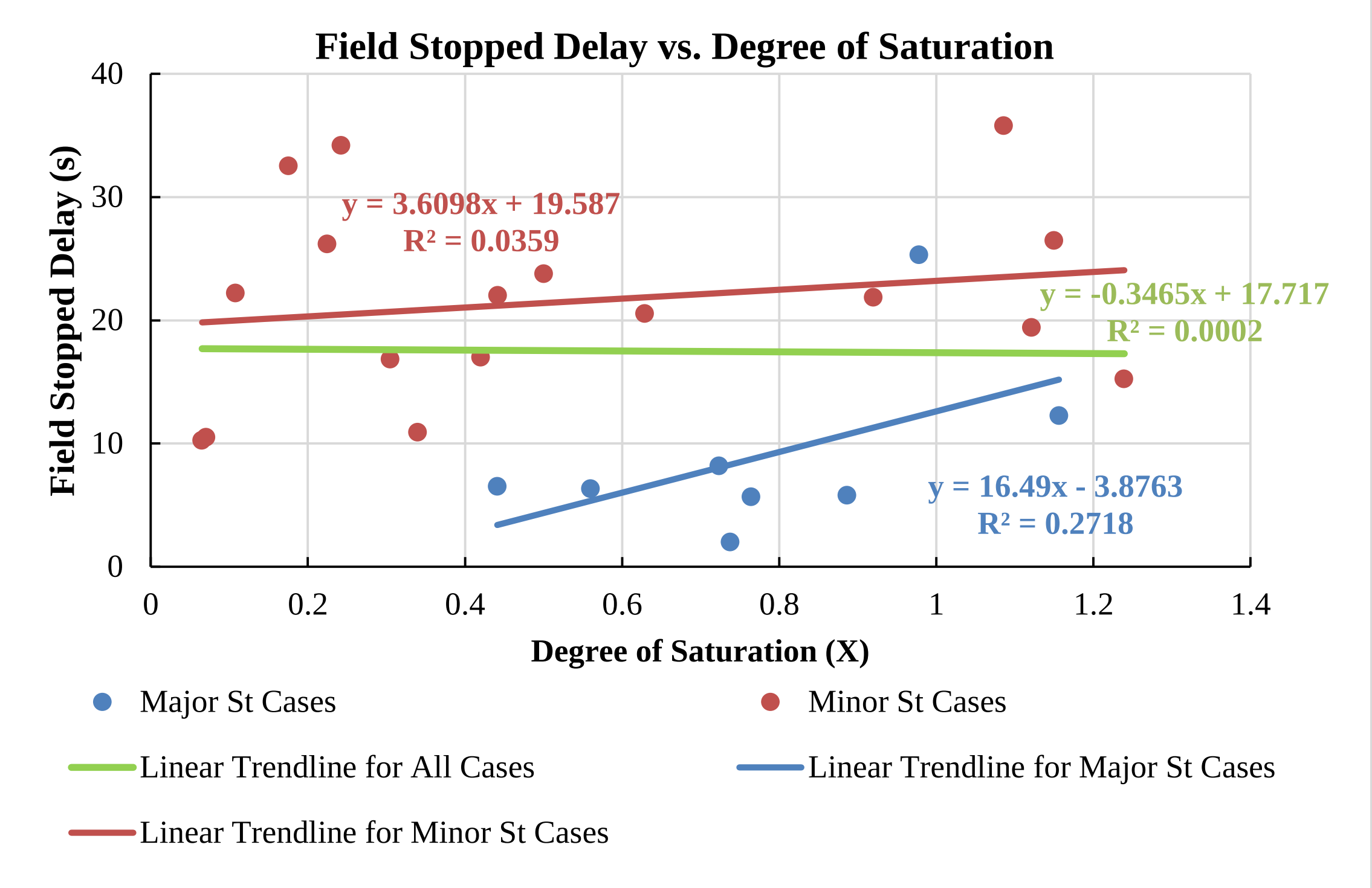

**FIGURE 8 Field Stopped Delay versus Degree of Saturation**

## CONCLUSIONS

Effects of different arrival types and degree of saturation (X) on HCM 2010 (and HCM 2016) procedure for the 50th percentile queue length at signalized intersections were analyzed (with a significance level of 0.1). Further analyses were conducted to establish the relationship between queue and delay with the degree of saturation when HCM yielded overestimated queue lengths.

A Chi-square test of goodness-of-fit was performed to examine the relationship among arrival type 4 and arrival types 2 and 3 combined, and the overestimated queue lengths on major and minor streets. The test results indicated that the arrival type did not have a significant effect in the distribution of overestimations among minor and major streets.

A Fisher's exact test performed to examine the relationship between overestimations and degree of saturation on the major and minor streets. The test results indicated that a significant correlation exists between degree of saturation (X) and overestimations on the minor and major streets. On major street as expected, the overestimations were within the range of v/c ratio higher than half (X>0.5). However, on the minor streets 59% of the overestimations were for X<0.49 and 23% were for oversaturated traffic conditions (X>1).

The relationship between amount of overestimations and degree of saturation should be established based on the numerical amount of overestimations versus X values; since the statistical comparison between the relative amount of overestimations and X values, resulted in a wrong idea of the real world condition.

Field queue length and stopped delay data of the cases with overestimations by HCM 2010 model are significantly correlated in all of the cases on major and minor streets. Furthermore, there is not a significant correlation between field stopped delay and X values of the cases with overestimation on minor and major streets; while there is a significant correlation between filed queue length and X values of all cases on major and minor streets.